\documentclass[fleqn,twoside]{article}
\usepackage{espcrc2}

\input BoxedEPS
\SetRokickiEPSFSpecial  
\HideDisplacementBoxes

\title{Continuum Extrapolation of Moments of Nucleon Quark Distributions in
Full QCD}

\author{LHPC and SESAM Collaborations:\\ P. Dreher\address[MIT]{%
        Center for Theoretical Physics,
        Massachusetts Institute of Technology,
        Cambridge, MA  02139, USA}%
\thanks{Support provided through the U.S. Department of Energy Cooperative
Agreement DE-FC02-94ER40818 and contracts DE-FG02-91ER40676,
DE-FG02-97ER41022, and DE-AC05-84ER40150},
R.\ Brower\address[BU]{Department of Physics,
                Boston University,  Boston, MA  02215, USA}$^\ast$,
S.\ Capitani\address[DESY]{John von Neumann Institut f\"ur Computing,
DESY,  D-15738  Zeuthen, Germany},
D.\ Dolgov\addressmark[MIT]$^\ast$,
R.\ Edwards\address[JLAB]{Thomas Jefferson National Accelerator Facility,      
Newport News, VA  23606, USA}$^\ast$,
N.\ Eicker\address[WUP]{Department of Physics,
                University of Wuppertal,  D-42097 Wuppertal, Germany},
U. M.\ Heller\address[FSU]{CSIT, Florida State University, Tallahassee, FL
32306-4120, USA}, Th.~Lippert\addressmark[WUP], 
J. W.\ Negele\addressmark[MIT]$^\ast$,
A.\ Pochinsky\addressmark[MIT]$^\ast$,
D. B.\ Renner\addressmark[MIT]$^\ast$,
K.\ Schilling\addressmark[WUP]}

\begin{document}

\begin{abstract}

Moments of light cone quark density, helicity, and transversity
distributions are calculated in unquenched lattice QCD at $\beta = 5.5$
and $\beta = 5.3$ using Wilson fermions on $ 16^3 \times 32 $ lattices.
These results are combined with earlier calculations at $\beta = 5.6$
using SESAM configurations to study the continuum limit.

\vspace{1pc}
\end{abstract}

\maketitle

\section{INTRODUCTION}

Calculation of moments of  light cone parton distributions using lattice
QCD provides
the only known way to calculate the quark and gluon structure of the
nucleon from
first principles.  Quenched calculations of flavor non-singlet moments of
parton
distributions yield well-known discrepancies with experiment
\cite{QCDSFquenched}, so
serious effort has been undertaken to calculate these moments in full
QCD\cite{horsley,LHPC}. To augment the calculations in ref.~\cite{LHPC} at
$\beta=5.6$,
this work reports calculations at $\beta=5.5$ and $\beta=5.3$ to study the
approach to
the continuum limit using Wilson fermions.

Moments of the quark density, helicity, and transversity distributions are
related to the following matrix elements of twist-2 operators:

\smallskip
\noindent $
{ \langle x^{n-1}\rangle_{ q_r } P_{\mu_1} \cdots P_{\mu_n}} $\\
\hspace*{.3in}$= \frac{1}{4} \langle PS | \left(\frac{i}{2}\right)^{n-1}
\bar{\psi}\gamma_{\{\mu_1}{\stackrel{\,\leftrightarrow}{D}}_{\mu_2} \cdots
{\stackrel{\,\leftrightarrow}{D}}_{\mu_n\}}\psi | PS\rangle$ \\

\vspace*{-.1in}\noindent
$ {\langle x^n \rangle_{ \Delta q_r} S_{\{\sigma}P_{\mu_1}\cdots
P_{\mu_n\}}} $\\
\hspace*{.3in}$= - \frac{n+1}{2}\langle PS | \left(\frac{i}{2}\right)^{n}
\bar{\psi}
\gamma_5\gamma_{\{\sigma}{\stackrel{\,\leftrightarrow}{D}}_{\mu_1} \cdots
{\stackrel{\,\leftrightarrow}{D}}_{\mu_n\}}\psi | PS\rangle $ \\

\vspace*{-.1in}\noindent
$ { \langle x^n \rangle_{ \delta q_r}
S_{[\mu}P_{\{\nu]}P_{\mu_1}\cdots P_{\mu_n\}}}  $\\
\hspace*{.3in}$= \frac{m_N}{2}\langle PS | \left(\frac{i}{2}\right)^{n}
\bar{\psi}
\gamma_5\sigma_{\mu\{\nu }{\stackrel{\,\leftrightarrow}{D}}_{\mu_1} \cdots
{\stackrel{\,\leftrightarrow}{D}}_{\mu_n\}}\psi | PS\rangle . $

\smallskip

We calculate four moments calculable using zero-momentum nucleon states:
the spin
averaged momentum fraction $\langle x\rangle_{ u-d }$,    the axial charge
$\langle 1
\rangle_{\Delta u - \Delta d}$, the longitudinal spin momentum fraction
$\langle x \rangle_{ \Delta u -  \Delta d}$, and the tensor charge
$ \langle 1 \rangle_{\delta u - \delta d}$.

\section{METHODOLOGY}

To facilitate direct comparison, we follow the same methodology as in
ref.~\cite{LHPC},
which used SESAM configurations at $\beta=5.6$ on a $16^3 \times 32$
lattice with
unimproved Wilson fermions. In particular, we calculated the ratios of
three-point to
two point functions using sequential Wuppertal smeared sources with the nucleon
projected to momentum zero and Dirichlet boundary conditions in the time
direction.
Renormalization constants to convert from lattice regularization to the
$\overline{MS}$
scheme at 4 GeV$^2$ were calculated perturbatively. 

At
$\beta=5.6$ the smearing was optimized to maximize the overlap of the
source with the nucleon ground state, yielding $N=50$ smearing steps, and the
source-sink separation 
was chosen to be 12 lattice steps to minimize the statistical errors
subject to the
requirement of a well-defined plateau in the three-point function.

At $\beta=5.5$ to keep the physical source size and the source-sink
distance
constant, we used 30 smearing steps and a separation of 10 lattice sites.
For $\kappa
= 0.1592$ we used 413 hybrid Monte Carlo configurations,  each separated by 20
trajectories, and at
$\kappa = 0.1596$, we used 243 configurations  separated
by 10 trajectories.

At $\beta=5.3$ we used 25 smearing steps and a source-sink separation
of 9 lattice
sites. At $\kappa = 0.1665$, we used 225 configurations separated by 15
trajectories, and at   
$\kappa = 0.1670$ we used 240 configurations  separated
by 10 trajectories.

\section{CHIRAL EXTRAPOLATION}

 \begin{table}[t!]
\caption{Moments of the quark distributions}
\label{table:3}
\newcommand{\m}{\hphantom{$-$}}
\newcommand{\cc}[1]{\multicolumn{1}{c}{#1}}
\renewcommand{\tabcolsep}{.25em} 
\renewcommand{\arraystretch}{1.2} 
\begin{tabular}{@{}cccccc}
\hline
\cc{$\beta$} & \cc{$\kappa$} & \cc{$\langle x\rangle _{u\!-\!d}$} & \cc{ $\!\langle 
1\rangle _{ \! \Delta{u} \!- \!\Delta{d}}$} & \cc{$\langle x\rangle _{ \! \Delta{ \!u} \!-
\!
\Delta{ \!d}}$} & \cc{$ \!\langle 1\rangle _{\delta{u}\!-\!\delta{d}}$} \\
\hline
5.3 & .1665 & .232(6) & 1.202(25) & .257(9) &  1.353(21) \\
5.3 & .1670 &.253(7) & 1.225(26) & .276(10) & 1.312(23)  \\
\hline
5.5 & .1592 & .230(3) & 1.180(12) & .262(4) & 1.273(9)
\\ 5.5 &.1596 & .238(5) &
1.140(17) & .265(5) & 1.277(13) \\
\hline
5.6 & .1560 & .247(9) & 1.190(16) & .297(6) & 1.325(13) \\
5.6 & .1565 & .242(9) & 1.223(20) & .298(7) & 1.287(18) \\
5.6 & .1570 & .263(13) & 1.094(25) & .283(9) & 1.261(26) \\
\hline
\end{tabular}\\[2pt]
\end{table}

Since the pion cloud contributes significantly to many of the operators
characterizing
form factors and moments of structure functions, it is reasonable to expect
large, qualitative differences between the behavior of matrix elements in
the heavy
quark regime where current lattice calculations are carried out and in the
light quark
regime pertinent to physical pions. A physical description of the
transition between
these two regimes is given by the following interpolation
formula~\cite{LHPC,detmold},
which incorporates the leading nonanalytic behavior in the pion mass
specified by
chiral perturbation theory:

\smallskip

\noindent $
{ \langle x ^n \rangle _{u-d}} $\\
\hspace*{.3in}$=  a_n \Bigl[ 1 -
{{(3 g_A^2+ 1) m_{\pi}^2} \over (4 \pi f_{\pi})^2} \ln \Bigl( {m_{\pi}^2\over
m_{\pi}^2 + \mu^2} \Bigr) \Bigr] + b_n m_{\pi}^2 .$ \\


The parameter $\mu$  specifies the physical scale of the three quark core
that serves
as the source term for the pion cloud, and a single value of this parameter
of the order
of 500 MeV resolves the discrepancy in the lowest three moments $\langle x
\rangle
_{u-d}$, $ \langle x ^2 \rangle _{u-d}$ ,  and $\langle x ^3 \rangle
_{u-d}$ as well as the
nucleon magnetic moment.  The best fit \cite{LHPC} to the $\beta =5.6$ data for 
$\langle
x \rangle _{u-d}$ is shown by the solid line in the upper panel of Figure~1. Of
particular importance for the present work is the fact that this interpolation
formula is nearly
constant in the regime in which all our lattice calculations are carried
out. Hence, even
though we only have data at two quark masses for $\beta = 5.5$  and  for
$\beta = 5.3$
we may study the approach to the continuum limit by determining the best least
squares fit of the parameter $a_1$ at each $\beta$. Analogous interpolation
formulae
for $\langle 1 \rangle_{\Delta u - \Delta d}$, $\langle x \rangle_{ \Delta
u -  \Delta d}$,
and  $ \langle 1 \rangle_{\delta u - \delta d}$ are presented in
ref.~\cite{adelaide} and
used to fit the data in the lower three panels of Figure~1.  Note that, in
contrast to the
spin averaged momentum fraction, the three spin-dependent moments are far less
sensitive to the pion mass because of strong cancellation between the the
nucleon and
delta contributions in chiral perturbation theory.

\begin{figure} [t!]
\BoxedEPSF{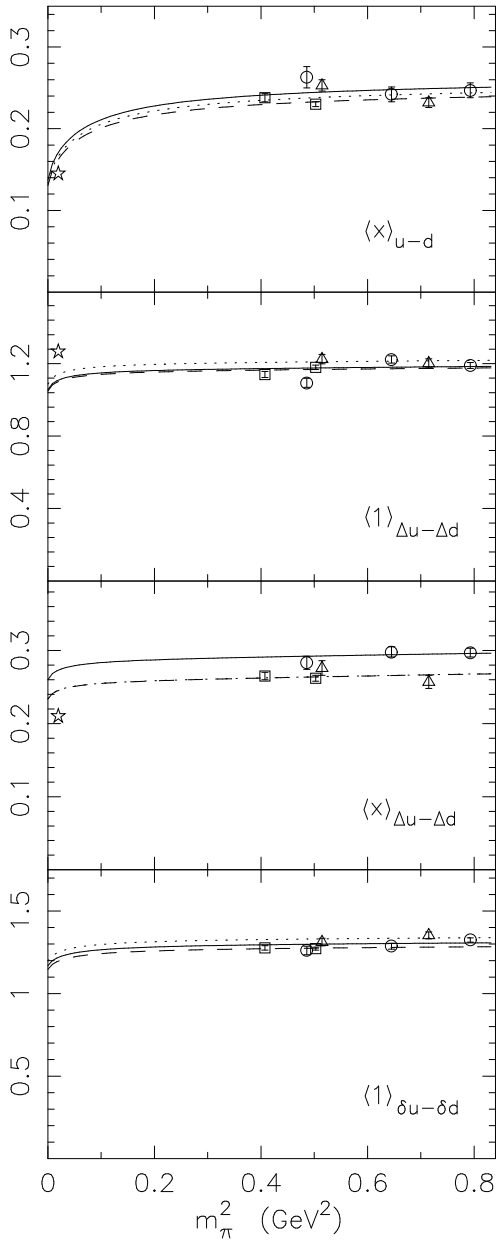 scaled 970}
\caption{Moments of quark distributions and interpolation formula fits at
$\beta=5.6$ (\emph{circles and solid lines}), 5.5 (\emph{squares and dashed
lines}), and 5.3 (\emph{triangles and dotted lines}). }

\end{figure}

\section{RESULTS}

The data are presented in Table 1. 
The results of least squares fits of the interpolation formulae  to the data at $\beta$
= 5.6, 5.5,  and 5.3 are shown in
Figure~1 by the solid, dashed, and dotted lines respectively. For each of
the four
operators, we see that the fractional shift in these lines is small as one
changes $\beta$. Especially since  the lattice spacing is quite large at
$\beta = 5.3$,
these results show that the ${\cal O} (a) $ corrections for unimproved
Wilson fermions
are small and benign for the operators considered in this work.  This
provides
solid evidence that the extrapolation to the continuum limit is reasonably
close to the
results at $\beta$ = 5.6. By ruling out significant $a$ dependence, this
result thus
further substantiates the physical argument that the origin of the discrepancy
between phenomenology and naive linear extrapolation of lattice moments of
parton
distributions must lie in the physics of the chiral extrapolation.

\subsection*{ACKNOWLEDGEMENT}

These calculations utilized some configurations at $\beta= 5.3$ from the HEMCGC
collaboration and at $\beta= 5.5$ from the SCRI group.  Additional production of 
hybrid Monte Carlo configurations and measurements of
operators were performed on the MIT alpha cluster, the JLab alpha cluster,
and the JLab QCDSP computer.

\end{document}